\documentclass[12pt]{article}

\usepackage{float}
\usepackage[dvipsnames]{xcolor}
\usepackage{hyperref}
\hypersetup{
    colorlinks,
    linkcolor=NavyBlue,
    citecolor=BrickRed,
    urlcolor=MidnightBlue
}
\usepackage{graphicx}
\usepackage{nth}
\usepackage[nointegrals]{wasysym}
\usepackage{amsmath}
\usepackage{amssymb}
\usepackage{authblk}
\usepackage{physics}
\usepackage{float}
\usepackage{nameref}

\usepackage[separate-uncertainty=true]{siunitx}
\usepackage{caption}
\newcommand{\vect}[1]{\boldsymbol{#1}}
\newcommand{\diff}{\,\mathrm{d}}
\newcommand{\NA}{\mathrm{NA}}
\newcommand{\exc}{\mathrm{exc}}
\newcommand{\emi}{\mathrm{em}}

\newcommand{\Int}{\int\displaylimits}
\newcommand{\FT}{\mathcal{F}}

\newcommand{\xs}{\vect x_s}
\newcommand{\xd}{\vect x_d}

\DeclareMathOperator*{\argmax}{arg\,max}

\makeatletter
\newcommand*{\currentname}{\@currentlabelname}

\usepackage[numbered]{bookmark}

\makeatletter
\newcommand*{\sectionbookmark}[1][]{%
  \bookmark[%
    level = #1,%
    dest=\@currentHref%
  ]%
}
\makeatother

\begin{document}

\title{Reconstructing the Image Scanning Microscopy Dataset: an Inverse Problem}

\author[1]{Alessandro Zunino}
\author[1,2]{Marco Castello}
\author[1,*]{Giuseppe Vicidomini}

\affil[1]{Molecular Microscopy and Spectroscopy, Istituto Italiano di Tecnologia, Genoa, Italy}
\affil[2]{Nanoscopy, Istituto Italiano di Tecnologia, Genoa, Italy}
\affil[*]{Correspondence: \href{mailto:giuseppe.vicidomini@iit.it}{giuseppe.vicidomini@iit.it}}

\date{}

\maketitle

\begin{abstract}
\noindent Confocal laser-scanning microscopy (CLSM) is one of the most popular optical architectures for fluorescence imaging. In CLSM, a focused laser beam excites the fluorescence emission from a specific specimen position. Some actuators scan the probed region across the sample and a photodetector collects a single intensity value for each scan point, building a two-dimensional image pixel-by-pixel.
Recently, new fast single-photon array detectors have allowed the recording of a full bi-dimensional image of the probed region for each scan point, transforming CLSM into image scanning microscopy (ISM). This latter offers significant improvements over traditional imaging but requires an optimal processing tool to extract a super-resolved image from the four-dimensional dataset.
Here we describe the image formation process in ISM from a statistical point of view, and we use the Bayesian framework to formulate a multi-image deconvolution problem. Notably, the single-photon detector suffers exclusively from the photon shot noise, enabling the development of an effective likelihood model. We derive an iterative likelihood maximization algorithm and test it on experimental and simulated data.
Furthermore, we demonstrate that the ISM dataset is redundant, enabling the possibility of obtaining reconstruction sampled at twice the scanning step. Our results prove that in ISM, under appropriate conditions, the Nyquist-Shannon sampling criterium is effectively relaxed. This finding can be exploited to speed up the acquisition process by a factor of four, further improving the versatility of ISM systems.
\end{abstract}

%
%
%
%
%

\section{Introduction}

Laser Scanning Microscopy (LSM) is the optical architecture at the base of many imaging and spectroscopy techniques  widely used in the fields of material science \cite{Teng2020VisualizationTechnique}, biology \cite{Bayguinov2018ModernMicroscopy}, and medicine \cite{Halbhuber2003ModernMedicine}. The reason for the success of LSM-based imaging techniques is their high spatio-temporal resolution and their capability to provide quantitative information. The LSM concept consists in focusing and scanning a laser beam on a sample to record either the scattered or fluorescent light with a single-element detector. Therefore, a complete image is built pixel-by-pixel by arranging the recorded intensity values along the scanning pattern \cite{Sheppard1977ImageMicroscope}.

Among the LSM techniques, Confocal Laser Scanning Microscopy (CLSM) became especially popular, thanks to its ability of rejecting out-of-focus light and its superior spatial resolution. In detail, CLSM setups are designed to image the focal plane of the objective lens onto a circular pinhole, placed before the detector.
Thus, the pinhole acts as a spatial filter allowing most of the light coming from the focal plane to reach the detector, while blocking most of the out-of-focus light. The closer the pinhole, the higher the spatial resolution and the signal-to-background (SBR) ratio of the images \cite{Conchello2005OpticalMicroscopy}.
Notably, CLSM images could achieve a lateral resolution twice better than the optical diffraction limit in the extreme case of a point-like pinhole aperture \cite{Sheppard1982Super-resolutionMicroscopy, Bertero1989Super-resolutionCase, Sheppard2017}. Thus, theoretically, CLSM should be a super-resolution technique, but in a realistic scenario, the pinhole cannot be fully closed. Indeed, closing the pinhole reduces the overall amount of recorded photons, compromising the signal-to-noise ratio (SNR) of the acquired images and hindering the capability of CLSM to achieve super-resolution.

Recently, image scanning microscopy (ISM) transformed confocal microscopy into a practical super-resolution technique. Theoretically developed in the 80s \cite{Sheppard1988Super-resolutionImaging, Bertero1987Super-resolutionMicroscopy, Bertero1989Super-resolutionCase}, it has been experimentally realized for the first time in 2010 \cite{Muller2010ImageMicroscopy}. The core idea behind ISM is to replace the single-element detector with an array of detectors, each detector acting as in a closed-pinhole configuration. Thus, after a complete scan, each detector element generates a confocal image which represents the same sample from a slightly different point-of-view. Furthermore, each detector contributes to the photon detection efficiency, maximizing the SNR of the acquisition. As a result, the ISM microscope provides a multi-dimensional dataset, which can be intuitively seen as a collection of scanned images, as many as the elements of the detector array, almost identical in content, but different in SNR and each shifted from each other by a quantity named \textit{shift-vector}.
Thus, estimating and compensating for the shifts enables the summation of all the collected images, constructing an ISM image with sub-diffraction resolution and enhanced SNR. This process is known by the name of \textit{pixel reassignment} (PR) \cite{Sheppard2013SuperresolutionReassignment}.

In the first ISM implementations, the PR method is performed using theoretically calculated shift-vectors -- which are based mainly on the geometry of the detector array \cite{Muller2010ImageMicroscopy, Sheppard2013SuperresolutionReassignment}. The clear advantage of this approach is its simplicity which also enables real-time analogue implementations of ISM \cite{Roth2013OpticalOPRA, DeLuca2013Re-scanResolution, York2013InstantProcessing}. However, the theoretical models used for the shift-vectors estimations are typically too simple to take into account non-idealities, such as misalignments or optical aberrations. Recently, a more robust approach, called \textit{adaptive pixel reassignment} (APR), was developed. The APR method estimates the shift-vectors directly from the experimental imaging dataset, measuring the cross-correlation of the scanned images \cite{Castello2015, Castello2019AFLIM}. As the name suggests, this approach takes inherently into account non-idealities, enabling a better reconstruction. Moreover, it enables the generalization of the ISM concept to other LSM techniques, such as two-photon excitation (2PE) microscopy \cite{Koho2020Two-photonReconstruction} and stimulated emission depletion (STED) microscopy \cite{Tortarolo2022Focus-ISMMicroscopy}.

Despite its great usability and vast potentialities, the APR method cannot be considered the most rigorous approach for reconstructing the ISM image. Indeed, APR builds on the assumption that the scanned images of the ISM dataset are all identical, but shifted and intensity-rescaled. This hypothesis is only approximately true and its validity highly depends on the design choices of the ISM microscope. Indeed, for detectors with a size larger than about one Airy Unit (AU), the point spread function (PSF) of the scanned images can differ also in the shape. Similar effects can arise in presence of strong optical aberrations, which are not uncommon in the context of thick sample imaging.

In this work we review multi-image deconvolution as the most comprehensive approach for reconstructing the ISM image. Deconvolution is a well-known image processing technique \cite{Bertero2021IntroductionImaging}, that consists in inverting the image formation process in order to remove the noise and the diffraction blurring \cite{Kabanikhin2008, Sibarita2005DeconvolutionMicroscopy}. LSM images are mainly affected by shot-noise, due to discrete nature of light and the negligible read-out noise of modern single-photon detector. Thus, algorithms that model the image formation as a Poisson process are particularly suited for deconvolving LSM images. Among them, the most successful algorithm is known with the name of the two scientists that conceived it, namely Richardson \cite{Richardson1972Bayesian-BasedRestoration} and Lucy \cite{Lucy1974}. Multi-image -- or joint -- deconvolution is a generalization of the Richardson-Lucy (RL) algorithm that estimates the most likely sample that generated a set of images that differ in some sense \cite{Ingaramo2014Richardson-LucyStrengths}. This approach has already been explored in the context of structured illumination \cite{Strohl2015AData}, ISM \cite{Castello2019AFLIM}, STED \cite{Castello2014Multi-imagesMicroscopy}, 4Pi \cite{Vicidomini2010AutomaticPhase}, and light-sheet microscopy \cite{Preibisch2014EfficientDeconvolution}. In this manuscript, we provide a complete theoretical framework for modeling the image formation process in ISM and its inversion. In detail, we derive the multi-image deconvolution algorithm and we provide a proof of its main properties. Notably, we provide evidence that ISM enables to sample twice less without losing any information on the image, thanks to the redundancy between the images of the ISM dataset. We extend the multi-image deconvolution reconstruction to take into account such redundancy and resample the final reconstruction. We validate our findings using both numerical simulations and experimental images of fluorescent samples. Thus, we demonstrate that ISM combined with multi-image deconvolution enables faster acquisition without image degradation, potentially enabling sampling beyond the Nyquist limit.

\section{The image scanning microscope}

\begin{figure*}[t!]
    \centering
    \includegraphics[width=\textwidth]{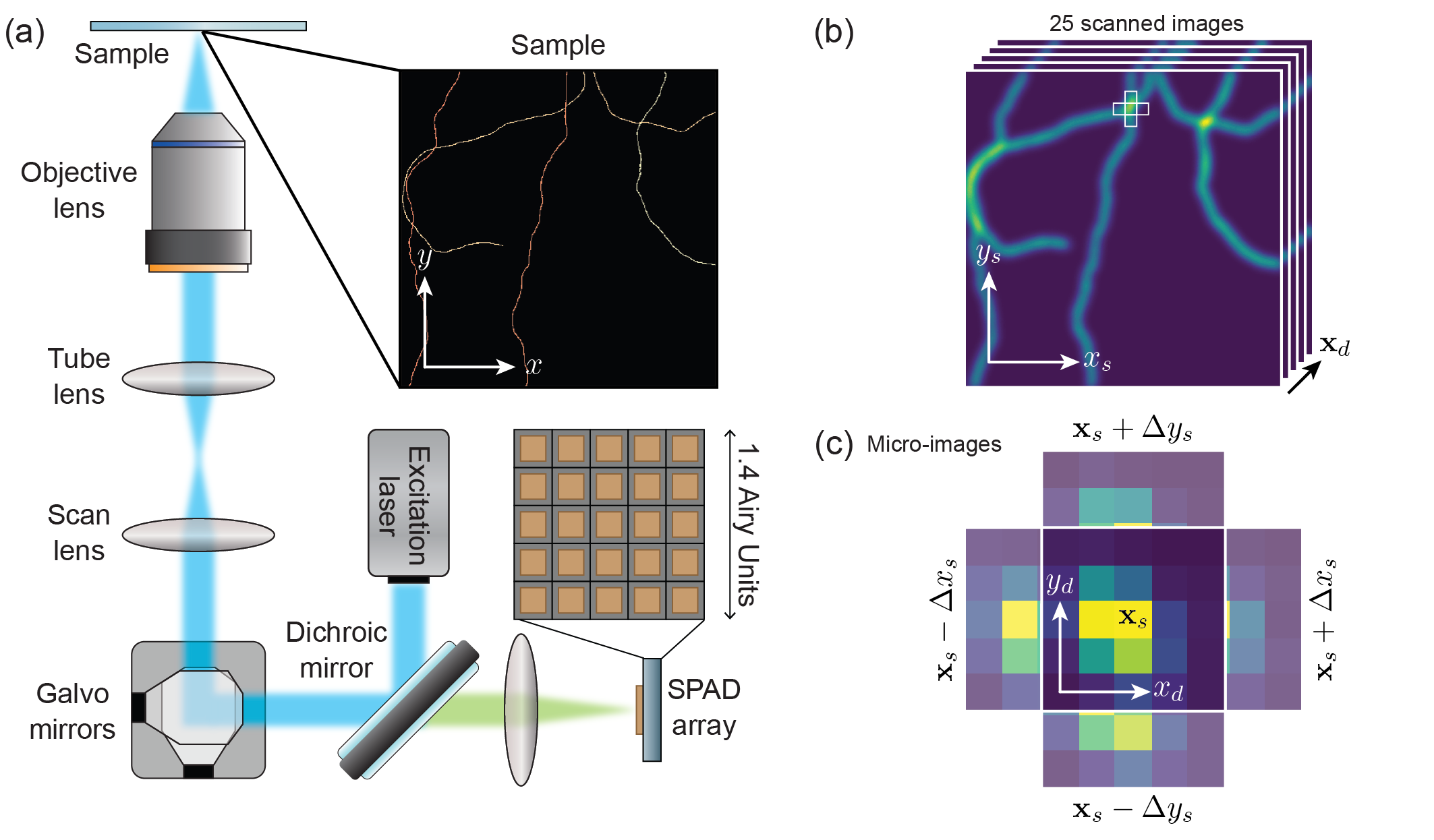}
    \caption{\textbf{Image Scanning Microscopy.} (a) A sketch of the laser scanning microscope equipped with a SPAD array detector. (b) The ISM dataset, seen as a set of scanned images as many as the number of elements of the detector array. (c) The ISM dataset, seen as a collection of micro-images, as many as the scan points. The depicted micro-images correspond to the scan points highlighted in (b) as white boxes.}
    \label{fig:1}
\end{figure*}

In this work, we performed fluorescence imaging using a custom image scanning microscope, sketched in Fig. \ref{fig:1}a. It consists of a standard laser scanning microscope, equipped with a \textit{single photon avalanche diode} (SPAD) array detector \cite{Buttafava2020SPAD-basedMicroscopy}. Such detector is composed by $5 \times 5$ SPAD elements, each acting as a small and displaced pinhole (0.28 A.U.). In other words, the ISM microscope works as 25 parallelized confocal microscopes, each recording an image of the sample from a slightly different point-of-view. After a complete scan, the collected fluorescence light is used to build 25 \textit{scanned images} $i(\xs | \xd )$, as depicted in Fig. \ref{fig:1}b. Here $\xs = (x_s, y_s)$ and $\xd = (x_d, y_d)$ are, respectively, the coordinates of the scan and the detector plane. Thus, the ISM dataset is four-dimensional, containing two extra spatial information with respect to a traditional CLSM image. Notably, taking in consideration the overall magnification of the microscope, both reference frames describe the same space -- namely the sample space -- whose coordinates are $\vect x = \xs - \xd$, hinting that the dataset contains some kind of redundancy. This latter can be better understood observing the ISM dataset from a different perspective. Indeed, it is possible to interpret the ISM dataset as a collection of \textit{micro-images} $i(\xd | \xs)$, namely small wide-field images of the illuminated region of the sample at the scan point $\xs$. Depending on the experimental values, namely the magnification of the microscope and the pixel size of the scanned images, neighbouring micro-images can partially overlap and share some information of the same region of the sample (see Fig. \ref{fig:1}c). The overlap ratio between adjacent micro-images can be quantified as

\begin{equation}
    \frac{D - M \Delta x_s}{D}
\end{equation}
where $D$ is the width the of the detector array, $M$ is the magnification of the imaging system, and $\Delta x_s$ is the scan step. For typical experimental parameters, such overlap can reach values above $90\%$. 

In the following sections we will show how to exploit such redundancy to either improve the SNR or the acquisition speed of the images. First, we devote this section to provide a theoretical description of the image formation process in ISM and its inversion using a maximum likelihood estimation approach.

\subsection{The forward problem}

\begin{figure*}[t!]
    \centering
    \includegraphics[width=\textwidth]{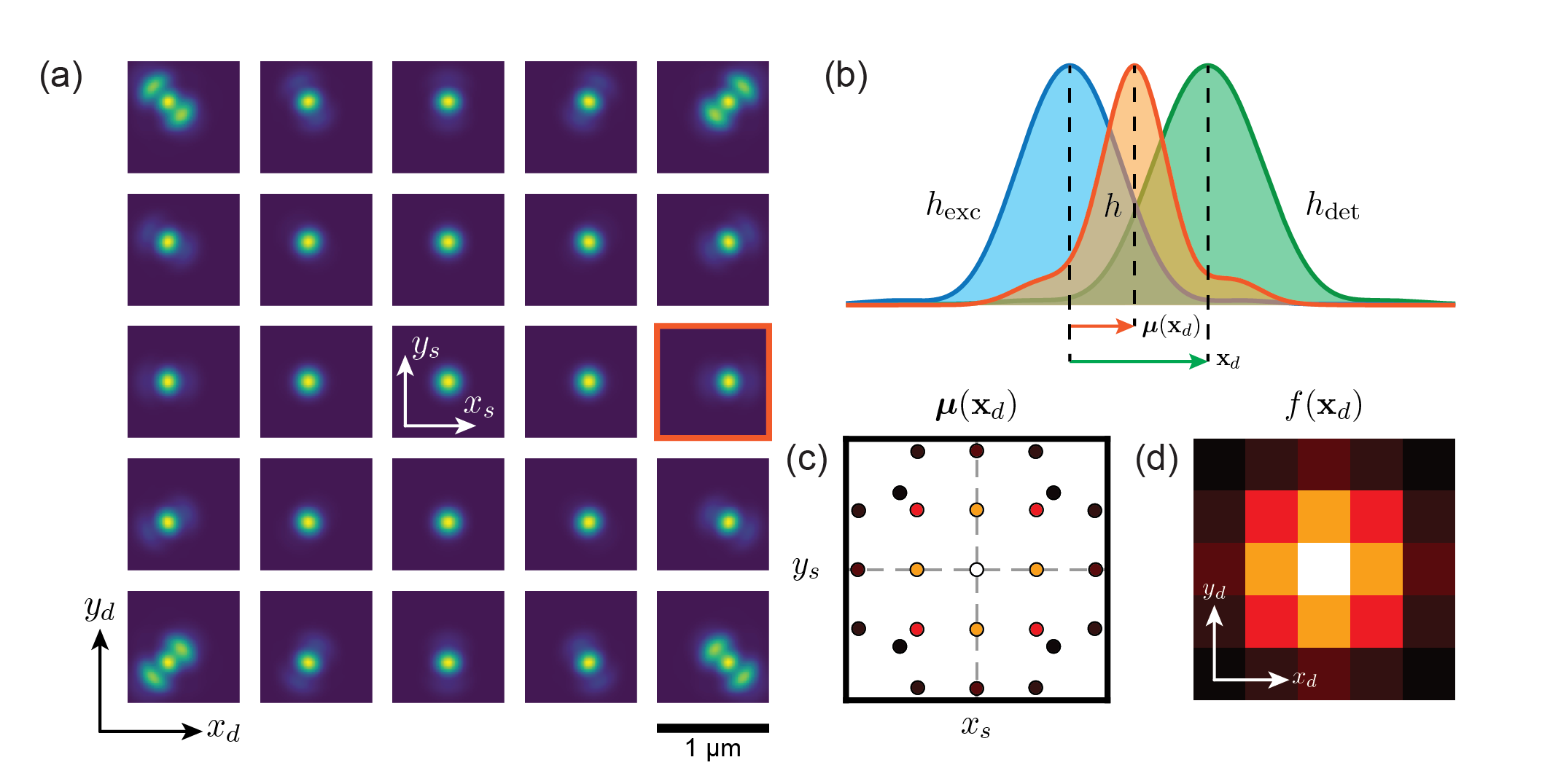}
    \caption{\textbf{ISM image formation} (a) Simulated images of $h(\xs|\xd)$, arranged as the corresponding detector element. (b) Line profile of the normalized PSFs corresponding to the detector elements highlighted with an orange box in (a). The blue, green, and orange curves represent, respectively, the excitation, detection, and complete PSF. The green and orange arrow are, respectively, the distance of the detector element from the optical axis and the shift-vector of the complete PSF. (c) Distribution of the shift-vectors of the whole ISM dataset, obtained with the APR algorithm. The plot is color-coded with the same colormap of the fingerprint, to denote the $\xd$ coordinate of each vector. (d) Fingerprint of the ISM dataset, obtained summing the dataset over the $\xs$ coordinate.}
    \label{fig:2}
\end{figure*}

\noindent The image scanning fluorescence microscope is a linear space-invariant system, whose output is a set of scanned images -- as many as the number $N_d$ of detector elements. The image formation process follows the same laws of CLSM, but we have to take into account the multiplicity of detectors and their relative position. The PSF of the scanned image generated by the detector element located at $\xd$ is

\begin{align}
    h(\vect x_s | \vect x_d) &= h_{\exc}(-\vect x_s) \cdot \qty[ h_{\emi}(\vect x_s) * p(\vect x_s - \vect x_d) ] = \nonumber \\
    & =  h_{\exc}(-\vect x_s) \cdot  h_{\det}(\xs - \xd)
    \label{eq:PSF_ISM}
\end{align}
where $p$ is the pinhole aperture. In our case, the pinhole represents the active area of the detector element. Through this work, we used squared active elements. The functions $h_{\exc}$ and $h_{\emi}$ are, respectively, the excitation and emission PSF of the microscope. We define the convolution product of $h_{\emi}$ with the pinhole $p$ as the detection PSF $h_{\det}$, namely an enlarged and shifted version of $h_{\emi}$. Fig. \ref{fig:2}a depicts the PSFs simulated for each detector element.

Notably, both $h_{\exc}$ and $h_{\det}$ are functions with a well-defined global maximum. Under the assumption that the PSFs have no other local maxima, also the complete PSF $h$ is a function with a single peak somewhere in between $\vect 0$ (the maximum position of $h_{\exc}$) and $\xd$ (the maximum position of $h_{\det}$) (see Fig. \ref{fig:2}b). Thus, we can write

\begin{equation}
    h(\vect x_s | \vect x_d) \approx h[\vect x_s - \vect \mu (\xd)]
    \label{eq:shiftedPSF}
\end{equation}
The exact maximum position of $h$ is called \textit{shift-vector} $\vect \mu (\xd)$ and, in general, depends on both the excitation and detection parameters (see Fig. \ref{fig:2}c). Importantly, the PSF $h$ is also narrower than $h_{\exc}$ and $h_{\det}$, enabling sub-diffraction resolution.

The assumption described by Eq. \ref{eq:shiftedPSF} is the basis of the APR method. Indeed, if all the images are just shifted replicas, the simplest way to combine the signal coming from the detector is to estimate the shifts, register the images, and sum them. However, the functions $h_{\exc}$ and $h_{\det}$ typically present multiple maxima, due to the presence of diffraction rings. Thus, the PSFs in general differ not only in position but also in shape. The most distorted PSFs are those associated to the most peripheral elements of the detector array, where the effect of the diffraction rings dominates. In this case the complete PSF $h$ might present multiple maxima with comparable intensity, making the definition of shift-vector ill-posed. However, the peripheral elements carry also the least amount of signal, as can be understood from the following argument.

The PSFs can be usefully rewritten in their normalized form $\bar{h}$

\begin{equation}
    h(\vect x_s | \vect x_d)=f(\xd) \cdot \bar{h}(\xs | \xd) 
\end{equation}
where the normalization factor is
\begin{align}
    f(\xd) & = \int_{\mathbb{R}^2} h(\xs | \xd) \diff \xs = \int_{\mathbb{R}^2} h_{\exc}(-\vect x_s) \cdot  h_{\det}(\xs - \xd) \diff \xs = \nonumber\\
    & = (h_{\exc} \star h_{\det}) (\xd)
\end{align}
Namely, the $N_d$ PSFs are rescaled by the factor $f(\xd)$ -- that we call \textit{fingerprint} -- which corresponds to the cross-correlation between $h_{\exc}$ and $h_{\det}$  (see Fig. \ref{fig:2}d).
Thus, the brightness of the PSF decays moving away from the detector center. This phenomenon explains the success of APR in practical scenarios: Eq. \ref{eq:shiftedPSF} holds for the PSFs that carry most of the signal. Nonetheless, this work aims to exploit the complete signal collected by the sensor, taking into account all the information at our disposal.

To this end, we write the scanned-image generated by the detector element at position $\vect x_d$ as the convolution between the object $o(\vect x_s)$ and the PSF

\begin{equation}
    i(\vect x_s | \vect x_d) = o(\vect x_s) * h(\vect x_s | \vect x_d)
    \label{eq:img_formation}
\end{equation}
The above relation describes $N_d$ forward problems, sharing the same source term $o(\vect x_s)$, laying the foundation of our multi-image deconvolution approach.

Notably, Eq. \ref{eq:img_formation} provides also a different way to estimate the fingerprint that does not require a measurement of the PSFs. Indeed,

\begin{align}
    \int_{\mathbb{R}^2} i(\vect x_s | \vect x_d) \diff \xs = \int_{\mathbb{R}^2} o(\xs) \diff \xs \cdot \int_{\mathbb{R}^2}h(\xs | \xd) \diff \xs = 
    \alpha \cdot f(\xd)
\end{align}
where $\alpha$ is a multiplication factor that depends uniquely on the object.

\subsection{The inverse problem}

In this section, we develop the solution of the inverse problem following a maximum likelihood estimation approach. In particular, we assume that the signal is affected only by photon shot noise -- which follows a Poisson distribution -- and that all the pixels are statistically uncorrelated \cite{Bertero2009ImageGalaxies}. Importantly, both of these assumptions are in good agreement with the specifications of the SPAD array detector, which has a low dark count rate and negligible cross-talk between active elements \cite{Buttafava2020SPAD-basedMicroscopy}.
The resulting likelihood probability is

\begin{equation}
    \mathcal{P} \qty[ i(\vect x_s | \vect x_d) \mid o(\vect x_s) ] = \prod_{\vect x_d} \prod_{\vect x_s} \frac{ [o(\vect x_s) * h(\vect x_s | \vect x_d)]^{i(\vect x_s | \vect x_d)} \cdot e^{-o(\vect x_s) * h(\vect x_s | \vect x_d)}}{i(\vect x_s | \vect x_d) !}
\end{equation}
where the product is understood to be a product integral.
The corresponding log-likelihood is
\begin{align}
    &\mathcal{L} \qty[ o(\vect x_s) ] = -\ln \qty{ \mathcal{P} \qty[ i(\vect x_s | \vect x_d) \mid o(\vect x_s) ] } = \\
    & = \iint \qty( o(\vect x_s) * h(\vect x_s | \vect x_d) - i(\vect x_s | \vect x_d) \cdot \ln \qty[ o(\vect x_s) * h(\vect x_s | \vect x_d) ] +\ln \qty[ i(\vect x_s | \vect x_d) !] ) \diff \vect x_s \diff \vect x_d \nonumber
\end{align}
Neglecting the additive terms that do not depend on the object, we find the following functional to be minimized
\begin{align}
    \mathcal{L} \qty[ o(\vect x_s) ] = \int \ell\qty[ o(\vect x_s) \mid \vect x_d ]  \diff \vect x_d
\end{align}
where we defined
\begin{align}
    \ell\qty[ o(\vect x_s) \mid \vect x_d ] = \int \qty( o(\vect x_s) * h(\vect x_s | \vect x_d) - i(\vect x_s | \vect x_d) \cdot \ln \qty[ o(\vect x_s) * h(\vect x_s | \vect x_d) ] ) \diff \vect x_s
\end{align}
In order to minimize $\mathcal{L}$, a convex functional, we have to calculate its functional derivative with respect to the object $o$.
Exploiting the linearity of the differentiation operator, we can calculate the derivative of $\ell$. This latter is

\begin{equation}
    \frac{\delta \ell}{\delta o} = h(-\vect x_s | \vect x_d) * \qty(1 - \frac{ i(\vect x_s | \vect x_d) }{ o(\vect x_s) * h(\vect x_s | \vect x_d) })
\end{equation}
Therefore, the derivative of the likelihood reads

\begin{align}
    \frac{\delta \mathcal{L} }{\delta o} &= \int h(-\vect x_s | \vect x_d) * \qty(1 - \frac{ i(\vect x_s | \vect x_d) }{ o(\vect x_s) * h(\vect x_s | \vect x_d) }) \diff \vect x_d  = \\
    & = \iint h(-\vect x_s | \vect x_d) \diff \vect x_s \diff \vect x_d - \int h(-\vect x_s | \vect x_d) * \qty( \frac{ i(\vect x_s | \vect x_d) }{ o(\vect x_s) * h(\vect x_s | \vect x_d) } ) \diff \vect x_d \nonumber
\end{align}
Assuming the normalization $\iint h(\vect x_s | \vect x_d) \diff \vect x_s \diff \vect x_d = 1$, we finally have

\begin{align}
    \frac{\delta \mathcal{L} }{\delta o} &= 1 - \int h(-\vect x_s | \vect x_d) * \qty( \frac{ i(\vect x_s | \vect x_d) }{ o(\vect x_s) * h(\vect x_s | \vect x_d) } ) \diff \vect x_d
\end{align}
We use this result to build the following gradient-descent method to find the minimum of $\mathcal{L}$

\begin{equation}
    o_{k+1} = o_{k} - \gamma_k o_{k} \frac{\delta \mathcal{L} }{\delta o}\eval_k
\end{equation}
Imposing a unitary step size ($\gamma_k = 1 \; \forall k$), we have

\begin{equation}
    o_{k+1}(\vect x_s) =  o_{k}(\vect x_s) \int h(-\vect x_s | \vect x_d) * \qty( \frac{ i(\vect x_s | \vect x_d) }{ o(\vect x_s) * h(\vect x_s | \vect x_d) } )\diff \vect x_d
    \label{eq:multi-img-deconv}
\end{equation}
As a consequence of this choice, we have pixels initialized with positive values maintain the non-negativity and pixels initialized with zero value remain null.

Notably, we can consider more general image formation processes and treat them with the same approach. One of the most important application is the inclusion of a background term $b(\vect x_s | \vect x_d) $ in the forward model

\begin{equation}
    i(\vect x_s | \vect x_d) = o(\vect x_s) * h(\vect x_s | \vect x_d) + b(\vect x_s | \vect x_d) 
    \label{eq:bkg}
\end{equation}
Using the above equation, we can repeat the same calculation to obtain the following iterative solution of the inverse problem

\begin{equation}
    o_{k+1}(\vect x_s) =  o_{k}(\vect x_s) \int h(-\vect x_s | \vect x_d) * \qty( \frac{ i(\vect x_s | \vect x_d) }{ o(\vect x_s) * h(\vect x_s | \vect x_d) + b(\vect x_s | \vect x_d) } )\diff \vect x_d
    \label{eq:multi-img-deconv-bkg}
\end{equation}

\subsection{Conservation of the flux}

An important property of the multi-image deconvolution algorithm is that the quantity of photons in the restored image (i.e., estimated object) is unchanged after each iteration. In other words, the role of the deconvolution is to reassign the photons present in the original dataset to their most likely origin. The proof is the following

\begin{align}
    \int o_{k+1}(\vect x_s) \diff \vect x_s &=  \iint o_{k}(\vect x_s) h(-\vect x_s | \vect x_d) * \qty( \frac{ i(\vect x_s | \vect x_d) }{ o(\vect x_s) * h(\vect x_s | \vect x_d) } )\diff \vect x_s \diff \vect x_d = \nonumber \\
    & = \iiint o_{k}(\vect x_s) h(\vect x - \vect x_s | \vect x_d) \qty( \frac{ i(\vect x | \vect x_d) }{ o(\vect x) * h(\vect x | \vect x_d) } ) \diff \vect x  \diff \vect x_s \diff \vect x_d = \nonumber \\
    & = \iint o(\vect x) * h(\vect x | \vect x_d) \qty( \frac{ i(\vect x | \vect x_d) }{ o(\vect x) * h(\vect x | \vect x_d) } ) \diff \vect x  \diff \vect x_d = \nonumber \\
    &= \iint  i(\vect x_s | \vect x_d) \diff \vect x_s \diff \vect x_d
\end{align}
Therefore, the photon flux in the deconvolved reconstruction is unaltered -- within the numerical precision -- independently of the chosen number of iterations. Importantly, in presence of the background term in the forward model the conservation of the photon flux is lost.

\section{Results and discussion}

\begin{figure*}[t!]
    \centering
    \includegraphics[width=\textwidth]{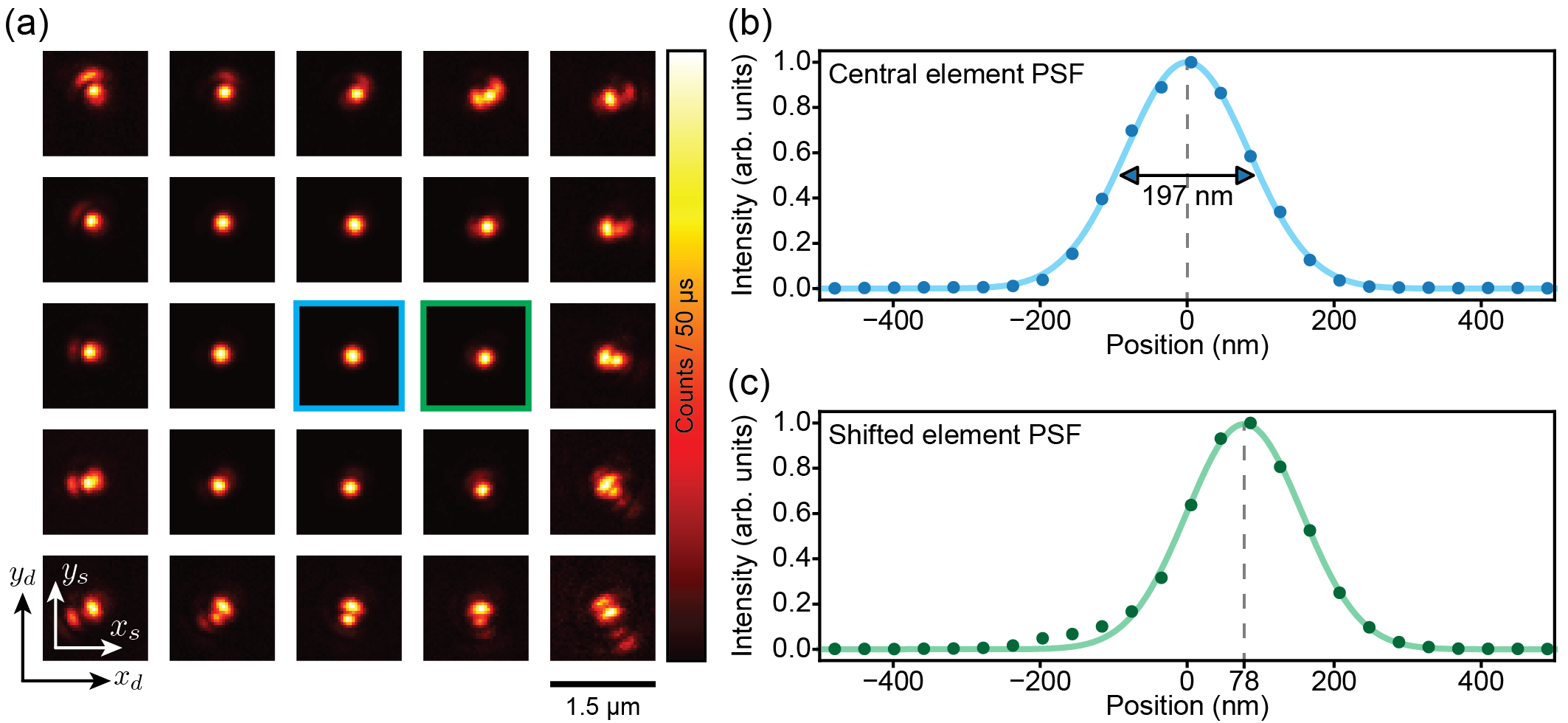}
    \caption{\textbf{Experimental Point Spread Functions.} (a) Images of the 25 PSFs obtained by scanning a sample of gold beads ($\diameter  = \SI{80}{nm}$) with a Gaussian beam at $\lambda = \SI{640}{nm}$. Each image is normalized to the maximum of itself. (b) Profile along the $x_s$-axis of the central element PSF, highlighted in (a) with a cyan frame. We report the FWHM, indicated with a double arrow. (c) Profile along the $x_s$-axis of a shifted element PSF, highlighted in (a) with a green frame. The $x_s$-projection of the corresponding shift-vector is indicated on the horizontal axis.}
    \label{fig:3}
\end{figure*}

In this section, we apply the multi-image deconvolution algorithm developed in the previous section on experimental data. We demonstrate its capability to reconstruct an image of the sample with higher spatial resolution and SNR. In order to apply the iterative formula shown in Eq. \ref{eq:multi-img-deconv}, we also require an estimation of the 25 point-spread functions (PSFs) of the ISM system -- one for each element of the detector array. The fidelity of the PSFs is key for achieving a satisfactory result, for this reasons different approaches has been proposed in literature. A possible approach is simulating the PSFs using a simple model, namely the PSFs are artificially calculated as Gaussian functions centered at the shift-vectors position and with a full width at half maximum (FWHM) equal to resolution estimated, for example, through the Fourier ring correlation (FRC) method \cite{Castello2015, Tortarolo2022Focus-ISMMicroscopy, Koho2019}. While being effective, this approach neglects the different PSF shapes and cannot exploit all the information contained in the dataset. A more exact approach would require a physics-based simulation, but this path involves the knowledge of several parameters which, in many conditions, are difficult to measure. Therefore, we choose an experimental approach and we measured the PSFs by acquiring the scanned images of a gold bead with a sub-diffraction size, as shown in Fig \ref{fig:3}a. This approach not only considers the physical parameters of the image formation process, but inherently takes into account non-idealities such as optical aberrations and misalignment. The choice of a scattering sample instead of a fluorescent sample inherently drops the effect of the Stokes shift -- which is typically negligible in terms of PSF structure -- but ensures high measurement stability and PSFs estimation with high SNR. Additionally, we analyzed the image of the central PSF, whose line profile is shown in Fig. \ref{fig:3}b. Fitting the profile to a Gaussian curve, we estimated the FWHM resolution of our system to be \SI{197.3(9)}{nm}, in perfect agreement with the Rayleigh prediction. Thus, we find that the Nyquist-Shannon sampling step for our microscope is $\sim \SI{100}{nm}$. We also fitted to a Gaussian curve the PSF of a neighbour detector element, as shown in Fig. \ref{fig:3}c. As expected from the theoretical model developed in the previous section, the peak position of the PSF is shifted. For this particular element, the shift is $\mu_{x_s} = \SI{77.9(8)}{nm}$.

\begin{figure*}[t!]
    \centering
    \includegraphics[width=\textwidth]{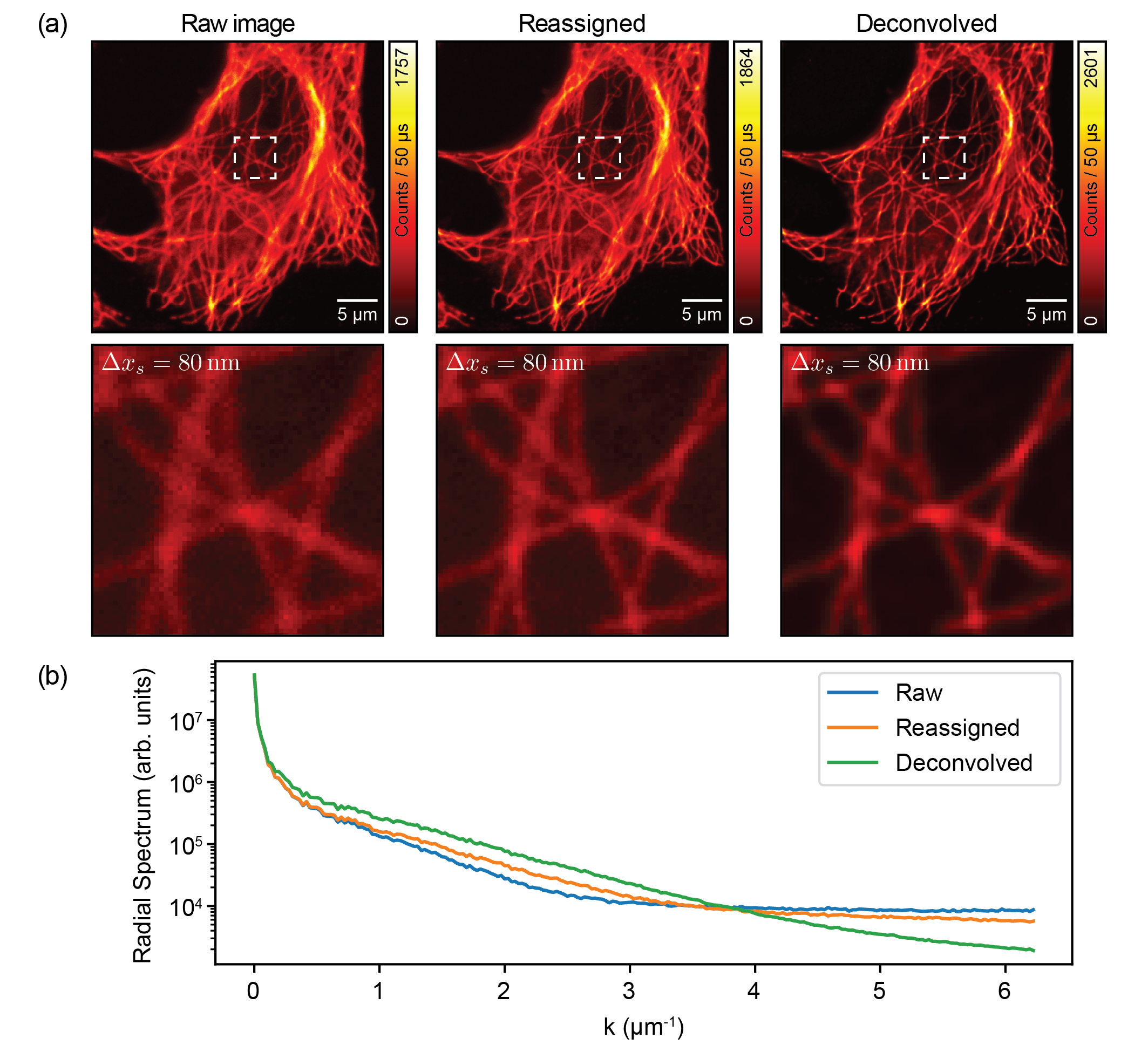}
    \caption{\textbf{Reconstruction of an experimental ISM dataset.} (a) Images obtained by summing the dataset (raw), applying the APR method, and applying the multi-image deconvolution using the PSFs shown in Fig. \ref{fig:3}a. The original dataset is sampled at \SI{80}{nm}, respecting the Nyquist-Shannon criterium. (b) Spectrum of the three images plotted as a function of the radial frequency $k$. The APR reconstruction (orange curve) shows an amplification of the high-frequency signal, due to the increase in resolution. The deconvolved reconstruction (green curve) shows an even greater enhancement of the high-frequency signal and a reduction of the noise level. The scale is shared between the three curves.}
    \label{fig:4}
\end{figure*}

Later, we acquired an ISM dataset of a fluorescent sample, namely a human cell with labeled tubulin network (see Fig. \ref{fig:4}a). We sampled the images at $\Delta x_s = \SI{80}{nm}$, thus respecting the Nyquist-Shannon criterium. We show the raw image, obtained by summing the photons collected by the whole detector, and the ISM reconstruction obtained using the APR method (see the Materials and Methods section for further details). Finally, we show the reconstruction obtained with the multi-image deconvolution after 5 iterations, calculated using the PSFs shown in Fig. \ref{fig:3}a. With APR, we got a reconstruction with the well-known enhancement of resolution and SNR. Although the APR method already provides a satisfactory result, the deconvolved reconstruction provides an even better improvement of resolution and SNR (see Fig. \ref{fig:4}b).

\begin{figure*}[t!]
    \centering
    \includegraphics[width=\textwidth]{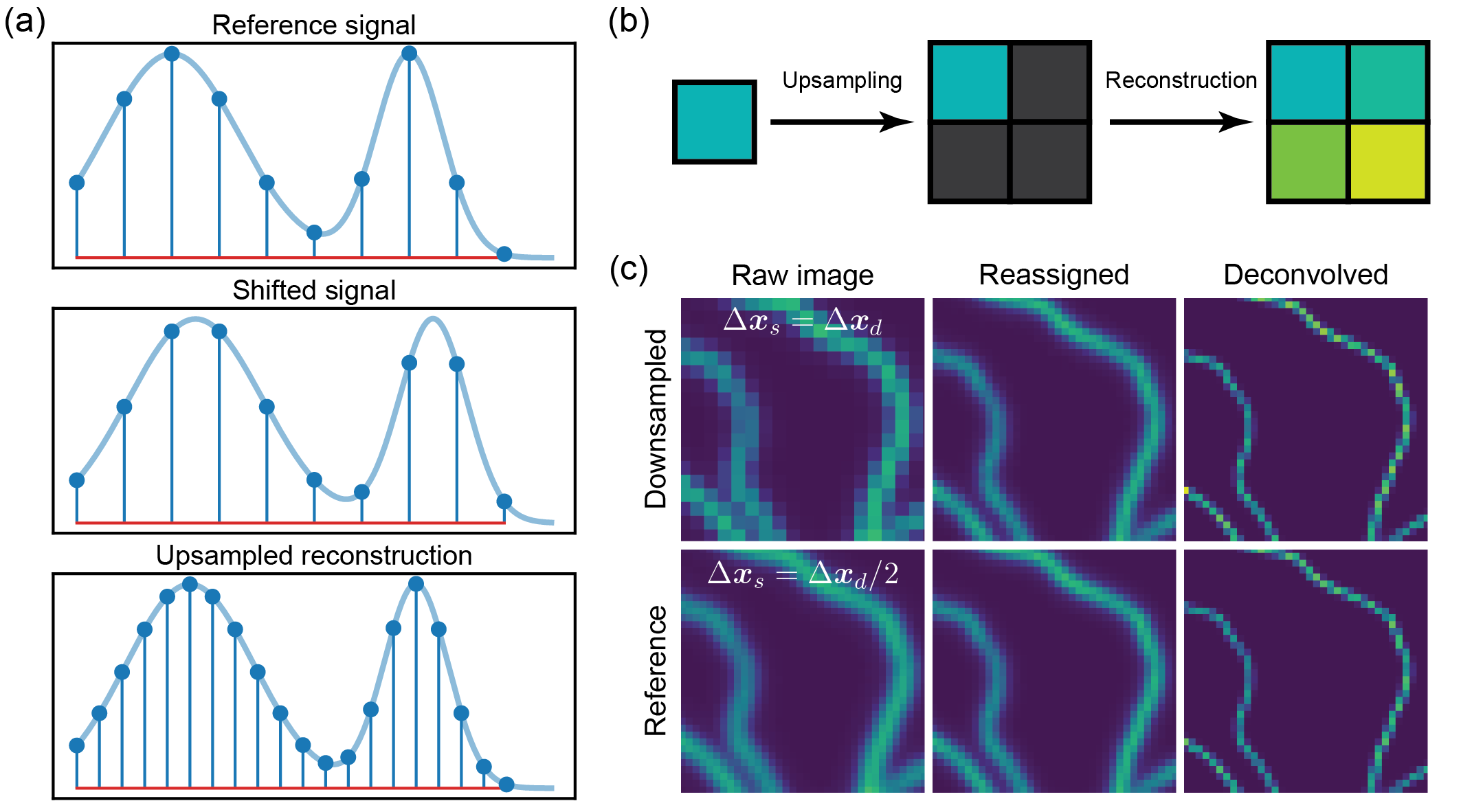}
    \caption{\textbf{Image upsampling.} (a) Working principle of the upsampling procedure. Sampling twice the same signal shifted by exactly half the sample step enables reconstructing the signal with doubled sampling rate. This process can be performed by alternating the samples from the original measurements. (b) Steps used to upsample the ISM reconstruction. Each pixel of each image is transformed into a square of four pixels, each zero-valued except for the top-left pixels that contains the original intensity value. Later, the reconstruction process (APR or deconvolution) fills the empty pixels. (c) Demonstration of the upsampling process on simulated data. The first row shows the raw downsampled dataset and the corresponding reconstructions. The second row shows the same reconstruction obtained from a dataset originally sampled at the target pixel size. We iterated the deconvolution algorithm 30 times in both cases.}
    \label{fig:5}
\end{figure*}

Nonetheless, the ISM reconstruction does not necessarily has to be used to increase the SNR. Instead, it can be used to improve the imaging speed by a factor of four without sacrificing the resolution. In order to understand the speed improvement, it is useful to recall that the images of the ISM dataset are approximately shifted replicas, as discussed in the previous section. The underlying principle is sketched in Fig. \ref{fig:5}a. During the acquisition, the same signal is recorded with the same sampling rate but with different shifts. If the sampling step is chosen to be exactly the double of the shift, then the two samplings can be combined into a single signal upsampled at twice the original frequency. In the context of ISM, the shift is the projection along each axis of the shift-vectors. The upsampling is performed by appending three zero-valued pixels for each original pixel of the images, as depicted in Fig. \ref{fig:5}b. In particular, we can rely only on the eight elements of the central $3 \times 3$ ring to compensate for the missing samples, since the most peripheral elements do not carry enough signal. Those images are used in the reconstruction process (either APR or deconvolution) to fill the empty pixels and to obtain an upsampled image. Thus, the upsampling condition for ISM is

\begin{equation}
    \Delta \xs  = 2 \cdot \vect \mu(\Delta \xd)
    \label{eq:condition1}
\end{equation}
where $\Delta \xs = \qty( \Delta x_s, \Delta y_s )$ is the scanning step, or pixel size, and $\Delta \xd = \qty( \Delta x_d, \Delta y_d )$ is the pitch of the detector array, projected onto the sample plane (i.e. rescaled by the magnification factor).

However, the value of the shift-vectors is typically not known prior a measurement. Thus, it is useful to provide a theoretical estimate. Indeed, the shift-vectors can be explicitly calculated in a simplified scenario, namely assuming Gaussian PSFs and no Stokes-shift. In this case, they assume the simple form $\vect \mu (\xd) = \xd / 2$. Furthermore, we assume the projections of the shift-vectors along the two axis to be identical, following the symmetry of the SPAD array detector. With these assumptions, the ISM sampling rule simplifies to

\begin{equation}
    \Delta x_s  = \Delta y_s = \Delta x_d = \Delta y_d 
    \label{eq:condition2}
\end{equation}
If such sampling condition is applied, then the images of the ISM dataset are shifted of exactly half pixel. Thus, it is possible to combine the information from the entire dataset to construct an ISM image with doubled sampling. This procedure is demonstrated with the simulations shown in Fig. \ref{fig:5}c. Indeed, if the pixel size is selected to respect Eq. \ref{eq:condition2}, the upsampled reconstructions are structurally indistinguishable from those obtained starting from a doubly sampled dataset.

\begin{figure*}[t!]
    \centering
    \includegraphics[width=\textwidth]{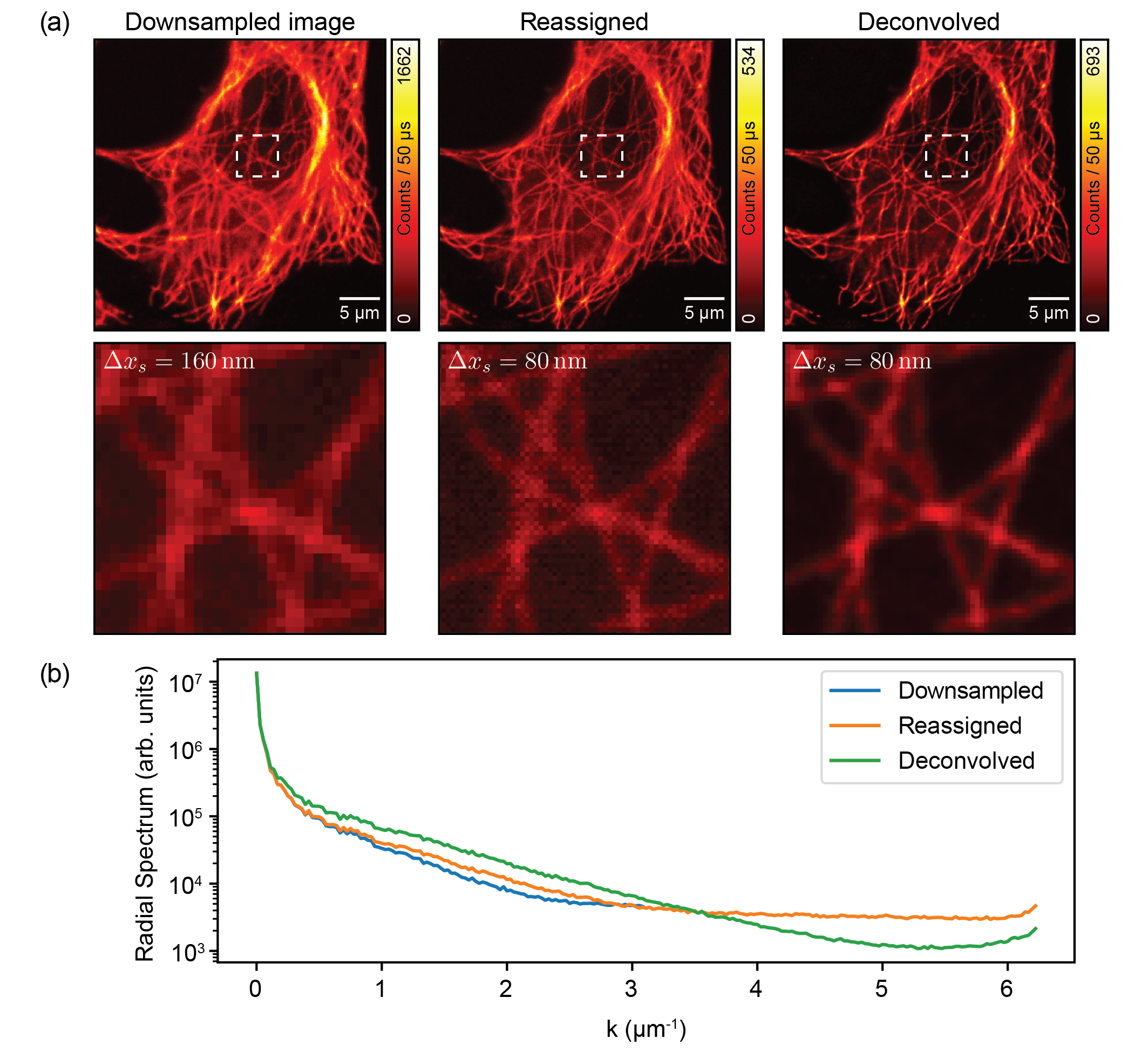}
    \caption{\textbf{Reconstruction and upsampling of a ISM dataset.} (a) Dataset obtained by downsampling the data shown in Fig. \ref{fig:4}a by a factor of two, thus being sampled at \SI{160}{nm}. Later, the dataset is upsampled using the procedure shown in Fig. \ref{fig:3}b and reconstructed using the APR method, and using multi-image deconvolution. The input data do not respect the Nyquist-Shannon criterium, but the final reconstructions do. (b) Spectrum of the three images plotted as a function of the radial frequency $k$. The APR reconstruction (orange curve) shows an amplification of the high-frequency signal and the deconvolved reconstruction (green curve) shows an even greater enhancement of the high-frequency signal and a reduction of the noise level, as in the reconstructions without upsampling. However, the reconstruction present some artifacts at the single-pixel scale which cause a rise of the radial spectrum close to the sampling frequency. The scale is shared between the three curves.}
    \label{fig:6}
\end{figure*}
These results imply that the Nyquist-Shannon criterion is relaxed for ISM by a factor of two. Indeed, it is possible to sample at half the Nyquist frequency as long as $\Delta \xd$ is designed to be equal or smaller to twice the Nyquist sampling step.

To prove this result on experimental data, we downsampled the dataset from \ref{fig:4}a of a factor of two. Thus, the pixel size of the downsampled dataset is \SI{160}{nm}, no longer respecting the Nyquist-Shannon criterium. Thanks to the procedure we just described, we obtained upsampled reconstructions, both using the APR method and the multi-image decovolution, shown in Fig. \ref{fig:6}a. However, it is difficult to achieve the condition described by Eq. \ref{eq:condition2} precisely. Small deviations from the ideal condition might cause the arise of artifacts in the reconstructions. This effect can be seen in the APR reconstruction, with clearly visible residues of the upsampling grid. Importantly, these artifacts are of the order of one pixel. In principle, they could be easily removed with a properly designed low-pass filtered, being spectrally well-separated from the signal (see Fig. \ref{fig:6}b). Nonetheless, we propose multi-image deconvolution as a superior reconstruction method for upsampled ISM images. Indeed, APR simply rigidly registers the 25 images by a fixed quantity. Instead, multi-image deconvolution builds the most likely structure that generated the images of the ISM dataset. Taking into account non only the shifts, but also the difference in the shape of the PSFs, we expect the multi-image deconvolution algorithm to provide a better reconstruction. Indeed, such expectation is confirmed by the result shown in Fig. \ref{fig:6}. The deconvolved reconstruction is structurally very similar to that presented in Fig. \ref{fig:4}a and the upsampling artifacts are greatly suppressed, but contains roughly one fourth of the original signal. Indeed, the redundant photons are exploited to interpolate the image, instead of being used to increase the SNR.

\section{Concluding remarks}

Reconstructing an ISM image is inherently an ill-posed problem that requires a properly designed strategy to be tackled. While APR is effective to produce images with enhanced resolution and SNR, it does not fully exploits the information encoded into the ISM dataset. As our results demonstrate, multi-image deconvolution enables superior reconstructions thanks to the inherent modeling of the image-formation process and noise statistics. Typically, the information embedded in the ISM dataset is partially redundant. The aforementioned reconstruction methods, namely pixel reassignment and deconvolution, normally use the excess of photons to enhance the signal of the image. However, our results demonstrate for the first time that the ISM redundancy can be used to upsample the final reconstruction, enabling an improvement of the imaging speed by a factor of four. Notably, the same advantage can be used to drastically reduce the sample exposure to the light, making ISM an even gentler technique. Indeed, if the magnified scanning step matches the pitch of the detector array, then the images of the dataset are shifted by half of the scanning step. Thus, we can perform a simple interpolation procedure: each pixel of the scanned images is alternated with an empty pixel. Later, the reconstruction algorithm fills the holes, generating an upsampled image. In principle, the upsampled reconstruction could be achieved with the APR method. However, the sampling condition needs to be matched exactly to have a good reconstruction, and small deviations from the ideality can generate visible artifacts in the image. We show that the multi-image deconvolution is a superior reconstruction method. Taking into account the exact PSFs of the system, the algorithm builds the statistically more likely upsampled image, with a significant reduction of artifacts.

Nonetheless, multi-image deconvolution requires a critical analysis. It is a well-known fact that deconvolution iterative methods are semi-convergent, meaning that the best reconstruction is achieved at a finite number of iterations \cite{Favati2014StoppingDeconvolution}. Indeed, iterating excessively the process could worsen the quality of the reconstruction, by amplifying the noise and generating artifacts. Therefore, in this work we used a conservative number of iterations (five) to deconvolve the experimental images. However, there is room for developing better approaches. A promising stopping criterium is found in the minimization the predictive risk estimation \cite{Massa2021PredictiveData}. The inclusion of an automated stopping criterium would make deconvolution a more robust approach for reconstructing the ISM image, ultimately becoming more user-friendly. In this work, we have not considered the inclusion of regularization rules. Indeed, it has been demonstrated that the implementation of sample-based priors, such as sparsity and continuity, helps improving the performances of deconvolution algorithms \cite{Zhao2022SparseMicroscopy,Vicidomini2009ApplicationMicroscopy}. Furthermore, the upsampled deconvolution could benefit from regularization rules favouring smooth solution, such as those penalizing the application of a differential operator on the estimated solution, in order to further reduce the presence of upsampling artifacts \cite{Sage2017DeconvolutionLab2:Microscopy}. Recently,
it has been proved that the combination of a deconvolution rule with a convolutional neural network provides fast and accurate reconstructions \cite{Li2022IncorporatingPerformance}. Thus, we believe that the method we proposed in this work could be further generalized by designing a neural network informed of the ISM image formation process.

In conclusion, this work provides the tools for an effective and reliable approach to ISM image processing and lays a solid foundation for the design of new methods. The existence of algorithms tailored for ISM will help spreading even further the adoption of detector arrays in laser scanning microscopes.

\section*{Materials and methods}
\sectionbookmark[1]{\currentname}

\subsection*{Custom setup}
\sectionbookmark[2]{\currentname}

\noindent For this work, we used a custom ISM setup equipped with a triggerable pulsed ($\sim 80$ ps pulse-width) diode laser (LDH-D-C-640, Picoquant) emitting at \SI{640}{nm} working as the excitation beam. This latter was deflected by two galvanometric scanning mirrors (6215HM40B, CT Cambridge Technology) and directed toward the objective lens (CFI Plan Apo VC 60$\times$, 1.4 NA, Oil, Nikon) by the same set of scan and tube lenses used in a commercial scanning microscope (Confocal C2, Nikon). The fluorescence light was collected by the same objective lens, descanned, and passed through the multi-band dichroic mirror as well as through a fluorescence band pass filter (685/70 nm, AHF Analysentechnik). A \SI{300}{mm} aspheric lens (Thorlabs) conjugates the fluorescence light to an image plane with a magnification of 300$\times$. A telescope system, built using two aspheric lenses of \SI{100}{mm} and \SI{150}{mm} focal length (Thorlabs), relays the image plane onto the detector plane with an additional magnification factor of $1.5\times$. The total magnification on the SPAD array plane is 450$\times$, thus the size of the SPAD array projected on the specimen is $\sim$ 1.4 A.U. (at the far-red emission wavelength, $\sim\SI{650}{nm}$). Every photon detected by any of the 25 elements of the SPAD array generates a signal that is delivered through a dedicated channel (one channel for each sensitive element of the detector) to an FPGA-based data-acquisition card. For this work we  used data-acquisition module from the PRISM-Light kit (Genoa Instruments). The kit includes a CMOS-based SPAD array detector with microlenses -- for improved collection efficiency.
We controlled the different components of the microscope and we recorded the data/images with our Python-based BrightEyes microscope software (BrightEyes-MCS), a custom program based on the Carma application \cite{Castello2017RemovalDetection}.
Specifically, the software also controls the entire microscope devices needed during the image acquisition, such as the galvanometric mirrors, the axial piezo scanner (PIFOC, Physik Instrumente), and the acousto-optic modulators (AOMs).

\subsection*{Sample preparation}
\sectionbookmark[2]{\currentname}

\paragraph{Gold beads.} A solution of gold beads (diameter: \SI{80}{nm}) in water was dropped onto a poly-l-lysine (Sigma)-coated glass coverslip. This latter was mounted with the same oil used as a medium for the objective lens (Nikon immersion oil).

\paragraph{HeLa cells.} The cells were fixed with ice methanol, 20 minutes at \SI{-20}{\celsius} and then washed three times for 15 minutes in PBS. After 1 hour at room temperature, the cells were treated in a solution of 3\% bovine serum albumin (BSA) and 0.1\% Triton in PBS (blocking buffer). The cells were then incubated with the monoclonal mouse anti-$\alpha$-tubulin antiserum (Sigma Aldrich) diluted in a blocking buffer (1:800) for 1 hour at room temperature. The $\alpha$-tubulin antibody was revealed using Abberior STAR Red goat anti-mouse (Abberior). The cells were rinsed three times in PBS for 5 minutes.

\subsection*{Numerical simulations}
\sectionbookmark[2]{\currentname}

\noindent We simulated the point-spread-function of the ISM system using the mathematical model presented in the manuscript. We simulated the detector array as a set of 25 squared pinholes arranged in a square matrix, using the geometry provided by the producer (pixel size: \SI{50}{\micro m}, pixel pitch: \SI{75}{\micro m}). We fixed the length of the detector, projected onto the sample plane, to 1.2 Airy unit (defined as the diameter of the Airy disc, $\SI{1}{AU} = 1.22\lambda_{\emi}/\NA$).  The excitation and emission wavelengths are, respectively, $\lambda_{\exc} = \SI{635}{nm}$ and $\lambda_{\emi} = \SI{660}{nm}$. We set the numerical aperture of the objective lens to $\NA = 1.4$ and the refractive index of the immersion oil to $n = 1.5$. To simulate the excitation and emission PSF, we performed a vectorial calculation using the pyFocus Python package \cite{Caprile2022PyFocusFoci}. The simulations are presented using the \textit{viridis} colormap, to make a clear distinction from the experimental results, shown using the \textit{hot} colormap.

\subsection*{Image reconstruction and analysis}
\sectionbookmark[2]{\currentname}

In this work, we performed the following processing of the ISM dataset.

\paragraph{Sum.} The raw dataset is quickly visualized by summing all the images of the ISM dataset:

\begin{equation}
    i_{\text{sum}}(\xs) = \sum_{\xd} i(\xs | \xd)
\end{equation}
The obtained image corresponds to a confocal image obtained with a pinhole as large as the detector array.

\paragraph{Adaptive pixel reassignment.} The adaptive pixel reassignment (APR) algorithm first calculates the correlograms, defined as the phase correlation between the images of the ISM dataset and the one generated by the central element, used as the reference.

\begin{equation}
    R(\vect x_s | \vect x_d) = \FT^{-1} \qty{ \frac{ \FT\qty{i(\vect x_s|\vect x_d)}   \cdot \overline{\FT\qty{i(\vect x_s|\vect 0)}} } { \qty| \FT\qty{i(\vect x_s|\vect x_d)} \cdot \overline{\FT\qty{i(\vect x_s|\vect 0)}} | } }
\end{equation}
Later, the algorithm finds the shift vectors as the position of maximum correlation

\begin{equation}
    \vect \mu( \vect x_d ) = \argmax_{ \vect x_s} \qty{ R(\vect x_s | \vect x_d) }
\end{equation}
and shifts each scanned image of the corresponding shift vector
\begin{equation}
    i(\vect x_s | \vect x_d) \xrightarrow[]{}  i\qty[\vect x_s + \vect \mu(\vect x_d) | \vect x_d]
\end{equation}
The final ISM image is finally calculated as

\begin{equation}
    i_{\text{ISM}}(\xs) = \sum_{\xd} i(\xs + \vect \mu(\vect x_d) | \xd)
\end{equation}

\paragraph{Multi-image deconvolution.} Multi-image deconvolution consists in the iterative reconstruction process described by equation \ref{eq:multi-img-deconv}. We implemented the following discretized version

\begin{equation}
    o_{k+1}(\vect x_s) =  o_{k}(\vect x_s) \sum_{\xd} h(-\vect x_s | \vect x_d) * \qty( \frac{ i(\vect x_s | \vect x_d) }{ o(\vect x_s) * h(\vect x_s | \vect x_d) } )
\end{equation}
The experimental results shown in this work have been obtained by fixing the maximum number of iterations to 5.

\paragraph{Downsampling and upsampling.} Being ISM a scanning technique, the proper way to downsample the images of the dataset is to drop one every two pixels for each axis. The resulting dataset is downsampled of a factor of 2 on the $\xs$ coordinate, thus containing one fourth of the pixels. The upsampling prior to image reconstruction is done by appending a zero-valued pixel to each pixel of the image for each axis.

\paragraph{PSF fitting.} We sliced the PSFs from Fig. \ref{fig:4} along the $x_s$-axis passing through the maximum position of the image of each analyzed PSF. Later, we fit the profile to the following Gaussian function

\begin{equation}
    I(x) = A \cdot \exp \qty[-\frac{1}{2}\qty(\frac{x-\mu}{\sigma})^2]
\end{equation}
We converted the standard deviation $\sigma$ to the FWHM using the following relation

\begin{equation}
\text { FWHM }=2 \sqrt{2 \ln 2} \sigma
\end{equation}
and measured the shift between the two PSFs as the difference between the fitted mean $\mu$ parameters.
The uncertainties are calculated as the square root of the diagonal elements of the covariance matrix returned by the fitting algorithm.

\paragraph{Radial spectrum calculation.} The spatial spectrum of an image $i(\xs)$ is calculated as its Fourier transform

\begin{equation}
    I(\vect k) = \FT\qty{ i(\xs) } 
\end{equation}
Where $\vect k = (k\cos\varphi, k\sin\varphi)$ is the spatial frequency vector, here written in polar coordinates.
Later, the spectrum is averaged with respect to the angular coordinate. The absolute value of the result is the radial spectrum

\begin{equation}
    S(k) = \qty| \frac{1}{2\pi} \Int_0^{2\pi} I(k, \varphi) \diff \varphi |
\end{equation}

\section*{Acknowledgments}
\sectionbookmark[1]{\currentname}

\noindent This research was supported by the European Research Council, BrightEyes, No. 818699 (G.V.). We thank Giorgio Tortarolo -- from École Polytechnique Fédérale de Lausanne -- for precious contributions in the development of the custom microscope. We thank Sabrina Zappone, Andrea Bucci, Francesco Fersini, Giacomo Garrè, Dr. Simonluca Piazza, Dr. Mattia Donato, Dr. Eli Slenders, Dr. Marcus Held, and Dr. Eleonora Perego -- from Istituto Italiano di Tecnologia -- for useful discussions.

\section*{Author contributions}
\sectionbookmark[1]{\currentname}

\noindent A.Z. and G.V. designed the study. A.Z., and G.V. designed and implemented the custom ISM system. M.C. contributed in the development of the deconvolution software. A.Z. and G.V. developed the analysis and simulation software. A.Z. developed the theory, performed the experiments and analyzed the data. A.Z. and G.V. wrote the manuscript. G.V. supervised the project. All authors discussed the results and commented on the manuscript.

\section*{Competing interest}
\sectionbookmark[1]{\currentname}

\noindent G.V. and M.C. have personal financial interest (co-founder) in Genoa Instruments. A.Z. declares no competing interests.

\bibliography{references.bib}
\bibliographystyle{unsrt}
\sectionbookmark[1]{\refname}

\end{document}